\documentclass[sigconf]{acmart}

\AtBeginDocument{%
  }

\setcopyright{none}
\acmConference[Preprint]{Preprint}{2025}{}%

\usepackage[most]{tcolorbox}
\usepackage{float}
\usepackage{stfloats}

\usepackage{xspace}
\tcbset{
  aibox/.style={
    top=10pt,
    colback=white,
    colframe=black,
    colbacktitle=black,
    enhanced,
    center,
    attach boxed title to top left={yshift=-0.1in,xshift=0.15in},
    boxed title style={boxrule=0pt,colframe=white,},
  }
}
\newtcolorbox{AIbox}[2][]{aibox,title=#2,#1}

\begin{document}


\title{SRLAgent: Enhancing Self-Regulated Learning Skills through Gamification and LLM Assistance}

\author{Wentao Ge}
\authornote{Both authors contributed equally to this research.}
%
\affiliation{%
  \institution{The Chinese University of Hong Kong, Shenzhen}
  \country{China}
}
\email{wentaoge@link.cuhk.edu.cn}

\author{Yuqing Sun}
\authornotemark[1]
\affiliation{%
  \institution{The Chinese University of Hong Kong, Shenzhen}
  \country{China}
}
\email{yuqingsun@link.cuhk.edu.cn}

\author{Ziyan Wang}
\affiliation{%
  \institution{The Chinese University of Hong Kong, Shenzhen}
  \country{China}
}
\email{ziyanwang@link.cuhk.edu.cn}

\author{Haoyue Zheng}
\affiliation{%
  \institution{The Chinese University of Hong Kong, Shenzhen}
  \country{China}
}
\email{haoyuezheng@link.cuhk.edu.cn}

\author{Weiyang He}
\affiliation{%
  \institution{The Chinese University of Hong Kong, Shenzhen}
  \country{China}
}
\email{weiyanghe@link.cuhk.edu.cn}

\author{PiaoHong Wang}
\affiliation{%
  \institution{City University of Hong Kong 
}
  \country{China}
}
\email{piaohwang2-c@my.cityu.edu.hk}

\author{Qianyu Zhu}
\affiliation{%
  \institution{The Chinese University of Hong Kong, Shenzhen}
  \country{China}
}
\email{zhuqianyu@cuhk.edu.cn}

\author{Benyou Wang}
\affiliation{%
  \institution{The Chinese University of Hong Kong, Shenzhen}
  \country{China}
}
\email{wangbenyou@cuhk.edu.cn}



\begin{abstract}

Self-regulated learning (SRL) is crucial for college students navigating increased academic demands and independence. Insufficient SRL skills can lead to disorganized study habits, low motivation, and poor time management, undermining learners’ ability to thrive in challenging environments. Through a formative study involving 59 college students, we identified key challenges students face in developing SRL skills, including difficulties with goal-setting, time management, and reflective learning. To address these challenges, we introduce SRLAgent, an LLM-assisted system that fosters SRL skills through gamification and adaptive support from large language models (LLMs). Grounded in Zimmerman’s three-phase SRL framework, SRLAgent enables students to engage in goal-setting, strategy execution, and self-reflection within an interactive game-based environment. The system offers real-time feedback and scaffolding powered by LLMs to support students’ independent study efforts. We evaluated SRLAgent using a between-subjects design, comparing it to a baseline system (SRL without Agent features) and a traditional multimedia learning condition. Results showed significant improvements in SRL skills within the SRLAgent group (\textit{p} < .001, Cohen’s \textit{d} = 0.234), and higher engagement compared to the baselines. This work highlights the value of embedding SRL scaffolding and real-time AI support within gamified environments, offering design implications for educational technologies that aim to promote deeper learning and metacognitive skill development.

\end{abstract}





\maketitle

\section{Introduction}
In higher education, students are expected to take ownership of their learning—setting goals, managing their progress, and reflecting on outcomes. 
This ability, known as self-regulated learning (SRL)\cite{zimmerman2002becoming, pintrich2004conceptual,harris1999programmatic,schraw2006promoting,schunk1996goal}, is a critical competency in college settings, where students face significantly greater demands for autonomy and self-direction than in high school.~\citep{appleby2006college,SONG2021learningpattern}. Unlike the structured and teacher-guided learning environment common in secondary education, college students must proactively plan, manage, and reflect upon their learning activities with minimal external supervision \cite{conley2008rethinking,venezia2013transition,crede2008study}. Prior research has demonstrated that effective SRL practices are strongly connected to students' academic achievement \cite{mega2014makes, dent2016relation, schunk2011handbook,broadbent2015self}, study satisfaction~\cite{liborius2019makes}, and long-term educational success\cite{theobald2021self}. Conversely, inadequate SRL skills among college students frequently lead to disorganized study habits, reduced intrinsic motivation, poor time management, lower academic achievement overall ~\cite{crede2008study,mega2014makes,zusho2017toward,dent2016relation,schunk2011handbook,dunlosky2012overconfidence}, and addictive behaviours~\cite{wang2020relationships}.
Cultivating SRL skills is therefore a critical priority in supporting college students’ academic development
 \cite{crede2008study,zusho2017toward,broadbent2015self,broadbent2017comparing}.

Previous research has explored various interventions and tools aimed at enhancing college students' self-regulated learning (SRL) skills. Educational interventions such as structured training sessions, workshops, and courses explicitly targeting SRL strategies have demonstrated improvements in student self-regulation and academic outcomes \citep{bail2008effects,paris2003classroom}. Additionally, technology-driven solutions including mobile applications~\cite{broadbent2015self}, web-based platforms~\cite{bellhauser2016applying}, and AI-driven tutoring systems~\cite{ng2024empowering,hsu2024artificial,molenaar2022concept} have emerged to provide personalized feedback, task management, and performance analytics aimed at fostering SRL. However, despite these advances, existing interventions exhibit several limitations. Explicit instructional programs often lack sustained engagement or fail to effectively motivate students, leading to limited transfer and application of SRL strategies beyond initial training contexts \citep{jarvela2013new}. Similarly, technology-based platforms frequently provide generic feedback or overly rigid scaffolds, lacking the adaptability necessary to address individual learner differences and emerging needs during real-time learning activities. Furthermore, many existing solutions do not adequately integrate motivational and engagement elements, such as gamification features, which could enhance learners' intrinsic motivation and sustained participation \citep{landers2014developing,koivisto2019rise}. Consequently, there is a critical need for innovative approaches that combine personalized, adaptive feedback mechanisms with engaging, motivationally supportive elements to effectively cultivate college students' SRL skills.

Building on these limitations, we conducted a formative study to investigate the specific design needs and better understand the challenges college students encounter while developing SRL skills. A total of 59 undergraduate students from diverse majors participated in this study, which incorporated a mixed-methods approach involving semi-structured interviews and surveys.

Based on findings from the formative study, we developed SRLAgent, a Minecraft-based interactive system designed to foster students' SRL skills with large language model(LLM) assistance. Grounded in Zimmerman's widely accepted three-phase model of SRL \cite{zimmerman2002becoming}, SRLAgent guides students through goal-setting (forethought), strategic monitoring and active learning (performance), and reflective evaluation of their learning experiences (reflection). By embedding SRL strategies directly within engaging gameplay, SRLAgent aims to seamlessly integrate skill practice into students' authentic learning experiences.

To assess the effectiveness of our proposed system, we conducted a between-subjects study comparing SRLAgent with two baseline conditions: a baseline version of SRLAgent without SRL support features and traditional multimedia learning. The study results showed that SRLAgent effectively promotes learners’ abilities in forethought and performance stages of Zimmerman’s SRL model, highlighting the value of explicit goal-setting and real-time feedback mechanisms. Descriptive trends indicate that SRLAgent may positively influence learner engagement and trust in AI systems, suggesting promising directions for future research and design
improvements.

This paper contributes to the field of educational technology and human-computer interaction (HCI) as follows:
1) \textbf{Development of SRLAgent}, an LLM-assisted interactive system designed to foster students' SRL skills; 2) Implementation of a \textbf{between-subjects user study} to evaluate SRLAgent's effectiveness and usefulness; 3)\textbf{Design implications} derived from the design process and user studies, offering guidelines for designing interactive systems that leverage generative AI and gamification.

\section{Related Work}

\subsection{Self‑Regulated Learning Strategies in Higher Education}
Empirical research has consistently shown the positive effect of SRL on academic achievement~\cite{azevedo2004does,chen2014web,ergen2017effect,broadbent2015self,dent2016relation,mega2014makes,schunk2011handbook}. Particularly, SRL is widely regarded as a foundational competency in higher education, where students are expected to take increasing responsibility for their learning and external guidance is reduced compared to secondary education\cite{xu2023synthesizing,SONG2021learningpattern,conley2008rethinking,crede2008study}. To foster this shift toward autonomy, numerous tools have been developed. Previous research has demonstrated the effectiveness of explicit instructional methods, such as structured workshops~\cite{arredondo1994using}, training courses~\cite{bail2008effects,theobald2021self,Dignath2008,}, and classroom-based programs~\cite{paris2003classroom,leidinger2012training,kistner2010promotion}, designed to directly teach SRL strategies to students. Meanwhile, digital tools have been proposed—such as study planners, metacognitive prompts, and learning analytics dashboards—designed to support goal-setting, routine structuring, and progress monitoring \cite{xu2023synthesizing,alvarez2022tools,roll2015understanding}. However, many of these tools adopt a static, one-size-fits-all model, offering limited personalization and interactivity. This lack of adaptability can hinder their effectiveness in meeting the diverse needs of learners \cite{delen2014effects}.

Recent studies emphasize that technologies that actively engage students in the SRL process are more effective in enhancing academic performance and motivation \cite{xu2023synthesizing}. Rather than simply tracking behavior, effective SRL tools encourage self-generated thinking and goal-directed action, which are particularly impactful in postsecondary contexts \cite{rienties2019unpacking, bellhauser2016applying}. Despite this, many current tools still fall short in offering real-time feedback, sustained engagement, or adaptive scaffolding—limitations that contribute to low retention and limited long-term impact on learning behavior \cite{rienties2019unpacking,bellhauser2016applying}.

To address these limitations, researchers have explored the development of adaptive learning systems powered by artificial intelligence. These systems continuously monitor learner progress and dynamically tailor instructional strategies, thereby increasing both engagement and learning effectiveness \cite{molenaar2023measuring}.


\subsection{Gamification in Education}
Gamification refers to the integration of game-design elements and game mechanics into non-game contexts, such as education, to enhance motivation, engagement, and learning outcomes \citep{deterding2011game}. In recent years, gamification has gained considerable attention as educators and researchers explore innovative ways to motivate learners, sustain attention, and improve educational experiences~\cite{barata2013improving,zeybek2024gamification,majuri2018gamification,oliveira2023tailored}. Numerous studies have empirically investigated the effectiveness of gamification in educational contexts, with generally positive outcomes reported across various domains and learner populations. For instance, gamified platforms incorporating badges, points, leaderboards, and quests have been shown to significantly improve student motivation, participation, and academic performance compared to traditional non-gamified instructional methods \citep{hamari2014does,seaborn2015gamification}. In higher education specifically, gamified learning environments have demonstrated improvements in learners' intrinsic motivation, engagement, and retention of course content \citep{koivisto2019rise}.

In parallel, the gamification of learning environments has emerged as a promising strategy to enhance students’ motivation and engagement in SRL. SRL skills such as goal-setting, self-monitoring, and reflection can be facilitated through gamification features like clearly defined quests (goal-setting), progress bars or badges (self-monitoring), and leaderboards or achievement systems (self-reflection and feedback) \citep{goslen2024leveraging,kim2009not}. Studies show that students exposed to gamified learning environments are more likely to consistently apply SRL strategies, maintain higher levels of persistence, and experience greater enjoyment in their academic tasks \cite{goslen2024leveraging}. Moreover, gamified systems can be enhanced with data analytics and real-time feedback, allowing educators to provide personalized guidance based on learners’ in-game behaviors and performance trajectories. This multimodal approach not only enhances situational engagement but also facilitates the development of SRL skills by helping students reflect on and adjust their learning strategies \cite{molenaar2023measuring}.

Despite its demonstrated benefits, gamification faces several limitations and challenges in educational contexts. One primary concern is the potential for superficial engagement; learners may become more focused on obtaining rewards or achievements rather than developing a genuine interest in the learning content \citep{nicholson2015recipe}. Additionally, poorly designed gamified systems may lead to cognitive overload, distraction from core educational goals, or even demotivation if students perceive the gaming elements as irrelevant or overly simplistic \citep{seaborn2015gamification}. There is also a noted lack of personalized adaptation in many gamified systems, which fails to account for individual differences in learner motivation, abilities, and preferences, potentially reducing their effectiveness for diverse learner populations \citep{koivisto2019rise}.

In our work, we addressed the gap in current gamified SRL interventions by integrating LLM-assisted real-time feedback and adaptive scaffolding within a gamified environment, allowing for personalized and dynamic support. This combination enhances both the motivational and metacognitive aspects of self-regulated learning, offering a more engaging and individualized learning experience.

\subsection{Large Language Models (LLMs) for Learning Support}
Large Language Models (LLMs), such as GPT-series~\cite{openaiChatGPT}, Claude-series~\cite{anthropicClaude}, and Gemini-series~\cite{googleGemini}, have emerged as powerful tools in the field of education, offering new ways to support learning and enhance student outcomes~\cite{zhang2024simulating,wen2024ai,kasneci2023chatgpt,opara2023chatgpt}. Grounded in deep learning and natural language processing (NLP), LLMs can understand complex instructional prompts, generate coherent and contextually appropriate explanations, and simulate interactive dialogues that closely mimic human tutoring interactions\citep{brown2020language,baidoo2023education,kasneci2023chatgpt}. Such capabilities allow LLMs to serve effectively as intelligent assistants, providing learners with real-time guidance and just-in-time help tailored to their individual needs and contexts.

Empirical studies have begun exploring various practical educational applications of LLMs. For instance, LLM-powered conversational agents have been deployed as interactive tutors, offering personalized explanations and feedback to students in domains ranging from mathematics to language learning \citep{baidoo2023education,stamper2024enhancing,10.1145/3613905.3648628}. Additionally, these models have been integrated into intelligent tutoring systems to assist students in problem-solving, scaffolding their reasoning processes, and answering questions dynamically \citep{ng2024empowering,hsu2024artificial}.

The integration of Large Language Models (LLMs) into educational contexts presents significant opportunities for enhancing self-regulated learning (SRL) among students.  Studies suggest that LLMs can provide personalized learning suggestions, prompting students to reflect on their learning experiences and improve their metacognitive strategies~\cite{abd2023large}. This aligns with self-efficacy theories, where positive feedback and personalized strategies boost a learner's belief in their abilities~\cite{calafato2023charting}. Also, research illustrates that LLMs can facilitate adaptive learning environments that adjust to individual student needs, subsequently fostering deeper engagement with the material~\cite{oppenheimer2024you}. Interactive agent-based systems have been shown to increase motivation and study efficacy by creating personalized learning pathways that respect individual learning preferences~\citep{oppenheimer2024you}.

In conclusion, while LLMs have shown promising potential for enhancing students' SRL skills, their practical application, particularly in directly supporting the SRL process, has not yet been thoroughly investigated. Our work aims to design a system integrating adaptive, real-time LLM-driven guidance with gamification elements, effectively enhancing both student motivation and personalized learning support. Additionally, our system specifically focuses on scaffolding students' self-regulated learning skills, mitigating common challenges related to superficial engagement and generic, non-adaptive feedback.

\section{Formative Study}

To better inform the design of SRLAgent, we conducted a formative study to evaluate the academic problems that students encounter. The findings of this study will guide the system's design concepts.  We used a mixed-methods strategy to collect quantitative and qualitative data, using standardized questionnaires and semi-structured interviews respectively.  The questionnaire was distributed to students from various academic years. Semi-structured interviews were done only with freshmen to acquire a better understanding of their early academic experiences.

\subsection{Participants and Procedure}
Participants in this questionnaire were current university students.    A total of 59 students completed the questionnaire, spanning academic years from freshman to senior, with freshmen accounting for 32.2\% of the total.    The survey included demographic questions such as students' year of study, gender, and major.    The questionnaire addressed a variety of academic adjustment issues, including perceptions of curriculum and course content, workload and assessments, time management and self-regulation, communication with instructors, academic support systems, and perceived academic pressure. 

To complement the questionnaire findings, we conducted semi-structured interviews with three university freshmen. The interviews were held via Tencent Meeting and focused on students' study habits, time management strategies, self-regulation practices, academic challenges, and their perceptions of AI-assisted and game-based learning systems.  Each interview lasted approximately 20 minutes.

\subsection{Findings}
Our formative study revealed significant insights into students' academic adjustment challenges and their perspectives on learning support. 

\subsubsection{Questionnaire results}

The questionnaire collected responses from 59 students regarding their study habits, time management, adjustment ability, and perceived academic pressure. 

The result on students' academic habits showed numerous distinct tendencies.  In terms of time management, the majority (54.2\%) of students stated that they can only "basically manage" their study time with occasional adjustments, while 16.9\% admitted to having problems managing their time and frequently encountering assignment delays.  This shows that many students fail to manage their time efficiently.

In terms of study state adjustment, 49.2\% of students said they "generally need some time" to change, while 15.3\% found it "difficult" to do so.  Only 25.4\% of students considered it easy to adapt to changing learning tasks, demonstrating the overall difficulties in sustaining appropriate learning settings.

When it comes to study and rest arrangements, more than half (50.8\%) of students reported having inconsistent study and relaxation schedules, indicating a lack of well-defined daily routines to enable sustained academic engagement.

In terms of progress alignment with course schedules, more than half (52.5\%) of respondents reported that their learning progress was regularly inconsistent with course schedules, highlighting difficulties with pacing and self-monitoring during the semester.

Finally, concerning perceived academic pressure, a substantial proportion of students (39\%) reported feeling "a high level of pressure" that was difficult to endure, while 32.2\% felt "a moderate amount of pressure" that, to some extent, motivated their academic progress.

The questionnaire results highlighted that students primarily need improved study skills and better support mechanisms, which aligns with our research focus on self-regulated learning (SRL) as a means to enhance academic performance.

\subsubsection{Semi-structured Interviews}

The follow-up semi-structured interviews with three participants focused on learning plans, study habits, time management, learning challenges, and perspectives on traditional education versus AI-assisted and gamified learning approaches. For clarity in the following discussion, we refer to these interview participants as S1, S2, and S3.

\textbf{Learning Process and Course Content.}
Regarding typical assignments, S1 and S2 mentioned that courses generally require quizzes, reports and occasional group projects that demand collaborative knowledge application. Participants commonly reported difficulty identifying key concepts when first encountering new material, often only recognizing knowledge gaps during examinations. S1 specifically noted that self-directed exploration becomes challenging without basic conceptual understanding, while S3 found traditional instructional methods effective due to their similarity with high school learning approaches.

\textbf{Learning Habits and Planning.}
Most participants described struggling with procrastination and maintaining consistent study routines, with few using detailed planning schedules. Students typically adopted spontaneous approaches driven by immediate needs or deadlines. Most did not use planning tools to track learning progress and engaged in minimal post-examination reflection, citing delayed feedback and lack of detailed solution explanations as barriers. This reflects a broader pattern of reactive rather than proactive learning management.

\textbf{Learning Support.} When encountering difficulties, students typically approached instructors and tutors only for specific problems they couldn't resolve independently, valuing thorough explanations until complete understanding was achieved. All participants also reported using AI tools when facing academic challenges. For AI learning support, students expressed clear preference for systems that explain problem-solving processes rather than simply providing answers. They valued interactive, conversational interfaces with friendly engagement styles that resemble effective human teaching.




\subsection{Design Principles for SRLAgent}
Based on our findings, we developed design principles for \textit{SRLAgent} grounded in Zimmerman's cyclical model of SRL, which aims to provide scaffolding that gradually develops students' independent learning capabilities, ultimately supporting their successful adjustment to the academic demands of higher education.

\begin{figure*}[htbp] 
\centering
\includegraphics[width=\textwidth]{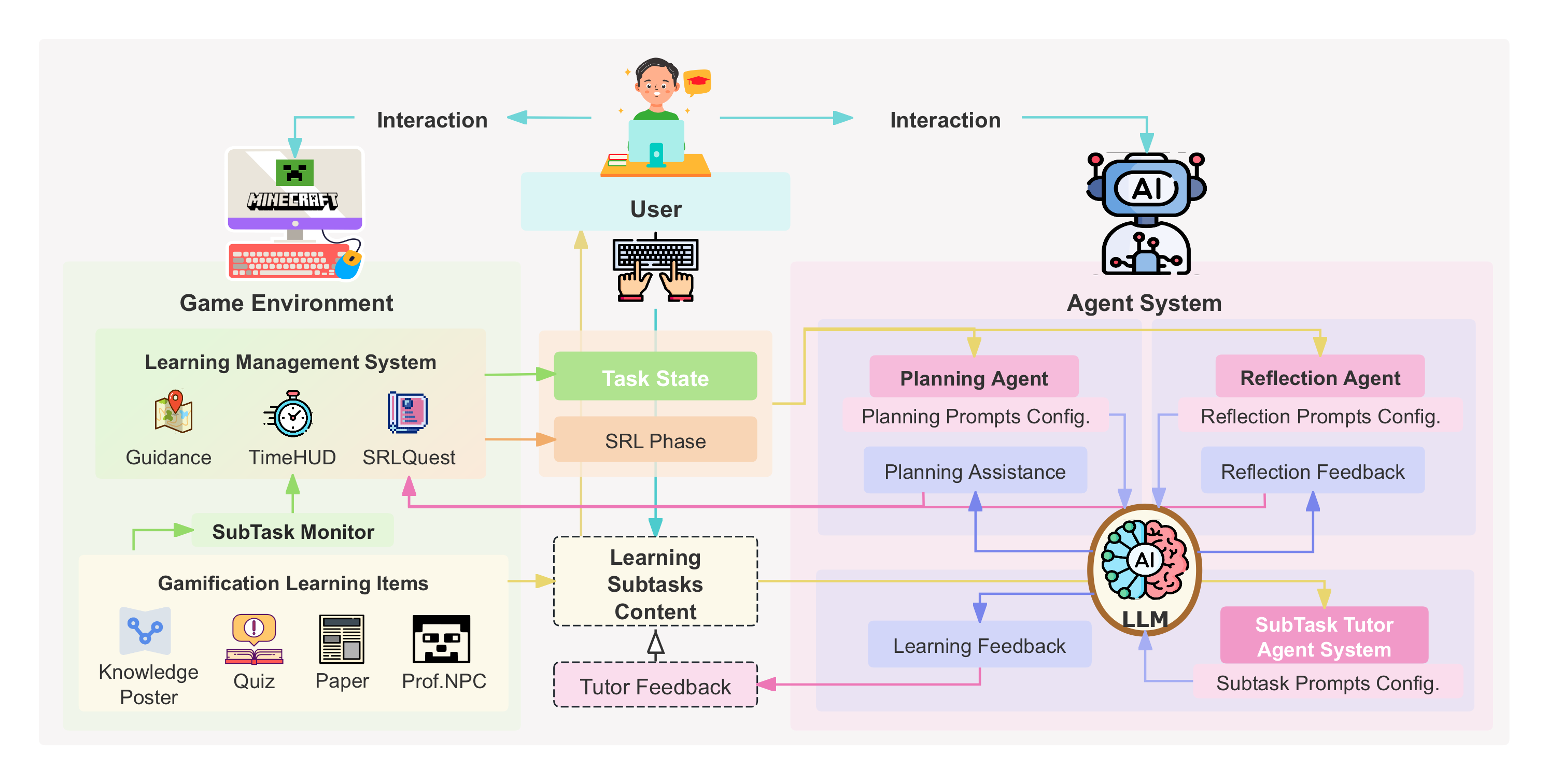}
\caption{SRLAgent System Architecture Overview}
\label{fig:system_architecture}
\end{figure*}

\subsubsection{Self-Regulated Learning for Academic Study Skills}
The first principle focuses on providing implicit SRL strategies to manage the learning process. This addresses the overarching management of the learning journey, supporting students in developing metacognitive awareness and skills. \textit{SRLAgent} will facilitate task planning and goal setting, time management assistance addressing the prevalent procrastination challenges, and comprehensive task summaries and feedback to promote reflection and continuous improvement.

\subsubsection{Real-time and Personalized Support in Learning Process}
The second principle emphasizes delivering real-time, learning task-specific suggestions during the performance phase of individual subtasks. Zimmerman's model presents self-regulated learning as three sequential phases. In our design, we aim to explore whether intelligent agents can effectively support self-regulated processes during the performance phase of various subtasks with real-time feedback. This design approach for \textit{SRLAgent} integrates immediate feedback mechanisms within active learning experiences for different subtask types. Our system provides timely feedback during active learning subtasks, including detailed explanations after quiz attempts, addressing students' desire for understanding principles rather than just receiving answers. For paper review and writing subtasks, the system offers real-time Q\&A and assistance with knowledge organization and writing guidance. Throughout the learning process, students can access specialized agent-based expertise tailored to specific subtask domains, enabling deeper conceptual discussions. Through this design, we explore how agents can scaffold self-regulated learning processes during active subtask performance.

\section{SRLAgent}

\begin{figure*}[htbp] 
\centering
\includegraphics[width=\textwidth]{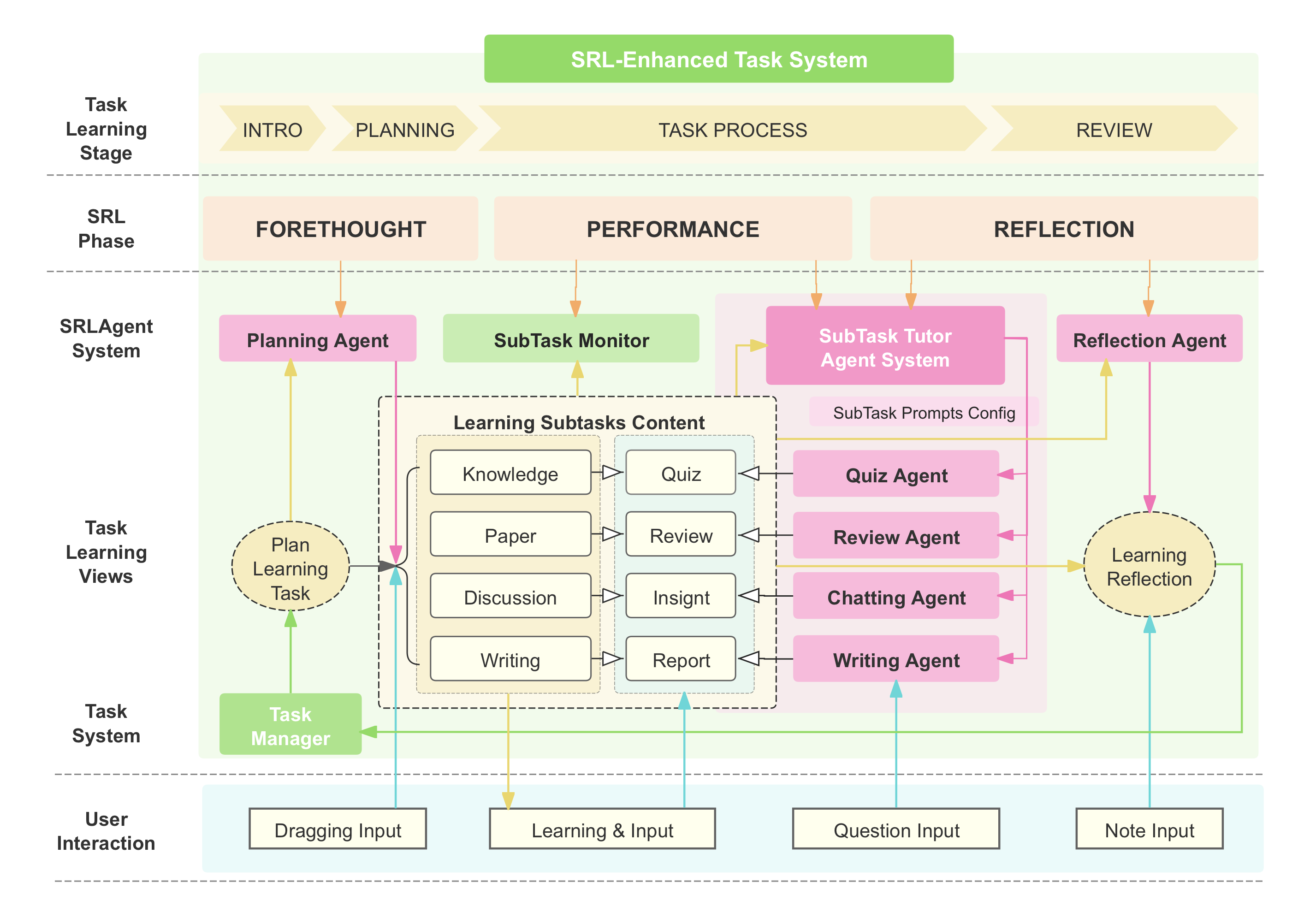}
\caption{SRLAgent framework integrating Zimmerman's three-phase SRL model with specific learning task stages.}
\label{fig:SRLAgent}
\end{figure*}

In this section, we present SRLAgent, the gamified learning environment built on the Minecraft platform that leverages large language models (LLMs) to foster SRL skills in college students. As shown in Figure~\ref{fig:SRLAgent}, SRLAgent integrates Zimmerman's three-phase SRL model (Forethought, Performance, and Reflection) with specific learning task stages to create a comprehensive framework that guides students through the entire self-regulated learning process. We designed SRLAgent's workflow and interaction based on the design principles obtained from our formative study.

The system consists of three key components: (1) an immersive 3D campus environment that provides a spatial context for learning activities through a detailed virtual representation of the university campus, (2) a flexible learning framework supporting customizable educational tasks through a hierarchical task management system with specialized learning activities organized in complementary pairs, and (3) a sophisticated LLM-driven agent system that delivers personalized guidance across all SRL phases through specialized agents for planning, monitoring, tutoring, and reflection. The implementation uses Minecraft's modding capabilities with Forge API, a structured MVC+System architecture for flexible content management, and an Agent Orchestration Framework that creates a bidirectional flow between the game environment and LLM-powered agents. Through specialized prompt templates aligned with each subtask content and SRL phase, the system delivers concise, contextually appropriate guidance that supports students' development of self-regulation skills while they engage with subject-specific learning content.

The following sections detail these system features and their implementation, followed by a case study demonstrating how the system supports students in developing self-regulated learning skills through this integrated approach.

\subsection{System Features}

\subsubsection{\textbf{3D Minecraft College Campus Environment [SF1]}}
The SRLAgent features a detailed virtual representation of the author's university campus within Minecraft. This immersive environment serves as the central setting for all educational experiences, creating a digital twin of the physical campus that includes key landmarks, academic buildings, residential colleges, and cultural elements. 

The virtual campus design supports spatial learning by connecting learning activities to a familiar academic context. Different campus locations host specific learning functions - libraries contain research materials, classrooms feature interactive knowledge displays, and meeting spaces enable NPC consultations - creating a coherent educational landscape that mirrors real university experiences.

\subsubsection{\textbf{Learning Skills Enhancement System [SF2]}}

\paragraph{Academic Content Integration}
SRLAgent integrates educational content directly into the virtual campus environment through interactive posters, digital libraries, and NPC knowledge bases. The system supports flexible content integration, allowing various academic subjects to be incorporated through a modular structure that separates content from delivery mechanisms. Educational materials are presented through spatially-distributed elements in the virtual environment, creating an immersive learning experience where knowledge discovery becomes part of exploration.

\paragraph{Task System}
SRLAgent employs a hierarchical task management system that orchestrates the learning experience across multiple levels. At the highest level, the system organizes the overall learning progression through four sequential stages as shown at the top of Figure~\ref{fig:SRLAgent}: \textit{Introduction} for initial orientation and context setting, \textit{Planning} for preparation and strategy development, \textit{Task Process} for core learning activities execution, and \textit{Review} for assessment and outcome evaluation.

The \textit{TaskManager} coordinates these stages and tracks major learning tasks assigned to the player. Each major task contains multiple subtasks with defined completion criteria, creating a structured progression path while maintaining flexibility for different learning contexts. As depicted in the center of Figure~\ref{fig:SRLAgent}, the system includes specific learning activities organized in complementary pairs: \textit{Knowledge} acquisition paired with \textit{Quiz} completion, \textit{Paper} reading paired with \textit{Review} creation, \textit{Discussion} participation paired with \textit{Insight} development, and \textit{Writing Goal} setting paired with \textit{Report} creation.

The \textit{TaskManager} is designed with a modular architecture that allows for enhancement through additional layers. Its core functionality tracks completion status across subtasks, synchronizes progression, and manages state transitions between activities to ensure a coherent learning flow. This design enables developers to rapidly implement new features that educational designers can subsequently utilize to create specialized learning experiences with different pedagogical approaches.

\paragraph{Study Skills Enhancement Layer}
Building upon the core TaskManager, our Task System can be extended to support various study skills frameworks through a Study Skills Enhancement layer with configurable views and tracking mechanisms. By adjusting task parameters, monitoring metrics, and feedback configurations, the system can be adapted to emphasize different educational approaches without modifying the underlying architecture. This extensibility allows educational designers to tailor the learning experience to specific pedagogical goals, whether focusing on time management, critical thinking, collaborative learning, or other study skills.

In this research, we specifically focused on enhancing Self-Regulated Learning skills. We implemented the Study Skills Enhancement layer with SRL stragegies and integrated a \textit{SRL-Enhanced Task System }that incorporates specialized agents corresponding to the phases of Zimmerman's SRL model. These SRL-focused components enhance the basic task system by providing real-time, customized feedback based on the player's progress and performance metrics. Each agent is specifically designed for particular learning contexts and subtasks, offering guidance tailored to both the educational content and the appropriate SRL strategies for that activity.

\subsubsection{\textbf{LLM-driven SRLAgent System [SF3]}}
Extending the capabilities of the Learning Skills Enhancement System [SF2], the SRLAgent system employs a sophisticated agent architecture that provides real-time, personalized support across all phases of the self-regulated learning process. This agent system is specifically customized to address both the educational content and the appropriate self-regulation strategies for each learning stage, enhancing the Task System with intelligent guidance mechanisms.

\textbf{Planning Agent}: Corresponds to the \textit{\textbf{FORETHOUGHT}} phase of the SRL cycle by guiding students in setting goals and developing strategies before beginning learning tasks. This agent helps learners establish clear objectives, select appropriate strategies, and develop a structured approach to their studies. It generates the "\textit{Plan Learning Task}" view shown in Figure~\ref{fig:SRLAgent}, which serves as both a planning document and a reference throughout the learning process.

\textbf{SubTask Monitor}: Operates during the \textit{\textbf{PERFORMANCE}} phase of the SRL cycle, tracking real-time performance across all learning subtasks and collecting data on time spent, completion rates, and quality indicators. This monitoring component serves as the foundation for the contextual awareness of other agents, enabling them to provide timely and relevant support based on the learner's current progress and challenges.

\textbf{SubTask Tutor Agent System}: Enhances the \textit{\textbf{PERFORMANCE}} phase through specialized agents tailored to specific learning activities. The \textit{Quiz Agent} offers adaptive support for assessment activities, identifying knowledge gaps and suggesting review strategies. The \textit{Review Agent} guides critical analysis and evaluation, helping students develop deeper understanding of materials. The \textit{Chatting Agent} facilitates productive discussions and insight development, promoting collaborative knowledge construction. The \textit{Writing Agent} provides structured support for report creation, assisting with organization, evidence use, and clarity. These specialized agents deliver immediate, contextual feedback aligned with both content mastery and self-regulation skill development.

\textbf{Reflection Agent}: Addresses the \textit{\textbf{REFLECTION}} phase of the SRL cycle by analyzing performance data and guiding self-evaluation, creating the "\textit{Learning Reflection}" view shown in Figure~\ref{fig:SRLAgent}. This agent helps students connect their strategies with outcomes, identify areas for improvement, and develop plans for applying insights to future learning activities.

Each agent integrates specialized prompt templates and context management systems to ensure interactions are appropriate, helpful, and aligned with both learning objectives and SRL principles. The system maintains conversation history and tracks progress data across multiple interactions, providing continuity throughout the learning experience and enabling agents to reference previous activities and outcomes. This coherent agent ecosystem creates a comprehensive support structure that guides students through the complete SRL cycle while adapting to their specific learning needs and subject matter requirements.

\subsection{Implementation}

\subsubsection{[SF1]\textbf{Minecraft Development and Modding}}
Corresponding to the 3D Minecraft College Campus Environment , our implementation involved:

\textbf{3D Campus Recreation}: Coordination with a specialized modeling team to accurately reproduce campus architecture, including precise spatial relationships and distinctive visual elements that ensure landmark recognition.

\textbf{Forge Mod Development}: Creation of a custom Minecraft mod using Forge API for Minecraft 1.18.2, written in Java. This extends Minecraft's base functionality to support educational interactions and SRL-specific mechanics.

\textbf{Interactive Elements}: Development of specialized blocks, items, and entities serving educational purposes, including interactive knowledge posters, paper displays, educational activity stations, and triggerable learning events.

\subsubsection{[SF2]\textbf{Learning System Architecture and Data Management}}
Supporting the Learning Skills Enhancement System , we implemented:

\textbf{MVC+System Architecture}: A structured code organization pattern that separates data models, visual components, control logic, and system services. This architecture supports the complex interactions between the TaskManager, learning subtasks, and user interface elements shown in Figure~\ref{fig:SRLAgent}.

\textbf{JSON Data Loading System}: An extensible data management framework that stores all educational content and task configurations externally. This includes task stage definitions, learning task specifications with dependencies, subtask completion criteria, educational content, and study skills enhancement settings. The separation of content from code enables rapid iteration and adaptation to different learning domains and study skills frameworks.
  
\textbf{Task Progression Framework}: A comprehensive tracking system that monitors student progression through interconnected learning subtasks, manages the game state and task availability based on player actions, synchronizes completed subtasks with parent tasks, and updates task views to reflect current learning status. Built on the modular architecture of the TaskManager, this framework provides the technical foundation for incorporating various study skills enhancements. In our implementation, we extended this framework to create the SRL-Enhanced Task System, which integrates specialized SRL agents with the core task management functionality to deliver context-aware self-regulation guidance throughout the learning process.

This implementation approach enables flexible content management while maintaining a consistent learning framework that can be enhanced with different pedagogical approaches.

\subsubsection{[SF3]\textbf{LLM-driven SRLAgent Implementation}}
To support the SRLAgent, we implemented:

\textbf{Agent Orchestration Framework}: The SRLAgent System serves as a coordination layer that facilitates the integration between Minecraft environment elements and AI-powered learning guidance as shown in Figure~\ref{fig:system_architecture}. Operating in concert with the SRL-Enhanced Task System, this framework utilizes task state information to determine when and how to activate different agents. As illustrated in Figure~\ref{fig:SRLAgent}, the orchestration aligns and activates agents with both Task Learning Stages and corresponding SRL Phases. This ensures that appropriate guidance is delivered at each point in the learning journey.

\textbf{LLM Toolchain}: This implicit middleware layer integrates LLM capabilities into our MVC+System architecture, treating AI services as system-level components. As shown in Figure~\ref{fig:system_architecture}, the toolchain manages the LLM API communications, context assembly from game state, and response processing. This approach creates a cohesive system where LLM capabilities are seamlessly embedded throughout the application, handling prompt construction, API calls, and response integration through a unified interface that other system components can access.

\textbf{SRL-Enhanced Prompt Templates}: Specialized system prompt designs for each agent type within the SRLAgent System shown in Figure~\ref{fig:SRLAgent}, with configurations tailored to their specific roles in the SRL process. These templates are organized hierarchically to align with the SRL-Enhanced Task System architecture:

\begin{itemize}
    \item \textbf{Forethought Phase Templates} (Planning Agent): 
    \begin{itemize}
        \item[] \textit{"You are an SRL expert. The player completes a series of tasks in the game and gets an Outcome at each step. Please summarize his task performance. Emphasis on using SRL skills to coordinate mission situations, and explicit SRL methods can be used, because the goal is to allow players to learn SRL skills. No more than 30 words. Please be concise, constructive, and clearly structured."}
    \end{itemize}

    \item \textbf{Performance Phase Templates} (SubTask Tutor Agent System):
    \begin{itemize}
        \item[] \textit{Writing Agent}: \textit{"You are an expert in report writing especially in University Education and Self-Regulated Learning(SRL) strategies. Your role is to assist the player in improving their SRL skills to achieve learning goals. The player now is in the Performance phase with subTask of writing a report according to Agent. Reply with no more than 30 words. You can answer by steps!"}
    \end{itemize}

    \item \textbf{Reflection Phase Templates} (Reflection Agent): 
    \begin{itemize}
        \item[] \textit{"You are an expert in University Education and Self-Regulated Learning (SRL) strategies. The player has completed a series of tasks in the game, each with its own outcome. The player is now in the Reflection phase of the SRL strategy. Please summarize their task performance based on the [Subtask Completion Content:] information you receive. No more than 30 words. Be concise, constructive, and clearly structured."}
    \end{itemize}
\end{itemize}

Each template incorporates phase-specific SRL principles and educational content relevant to the current learning activity. The concise format (limited to 30 words) ensures focused guidance while maintaining instructional quality across all phases of the learning process.

This implementation creates a bidirectional flow of information between the game environment and the agent system. Game elements provide context to the agents through the orchestration framework, while agent responses are delivered back to the user through appropriate interface elements like Task Guidance and SRLQuest, supporting the full SRL cycle depicted in Figure~\ref{fig:SRLAgent}.

\subsection{Case Study: Improving SRL Skills}
The SRLAgent system demonstrates how gamified environments integrated with LLM-powered agents can effectively support the development of self-regulated learning skills. Through the structured task progression and adaptive agent support, students learn not only subject matter content but also metacognitive strategies for planning, monitoring, and reflecting on their learning process.

To validate our approach, we customized an existing Hugging Face Agent tutorial as learning content within our SRLAgent system. We integrated this publicly available tutorial into our gamified learning system. First-year undergraduate students were invited to participate in testing sessions where they engaged with the Hugging Face Agent tutorial through our SRLAgent interface. 

The feedback collected from these sessions was overwhelmingly positive. Students reported that the system successfully enhanced their learning experience, with particular appreciation for the structured guidance provided by the SRL agents. Specifically, participants noted improvements in their metacognitive awareness and ability to plan learning activities strategically. 

\subsection{Example Use Case}

To illustrate how our system enhances self-regulated learning (SRL) skills among college students, we present a detailed scenario of a college freshman named Elvira. The Interfaces for this case are shown in the Appendix.

Elvira begins by opening her computer and launching the SRLAgent system within Minecraft. Upon entering SRLAgent, the system first helps her clearly understand her upcoming learning task by breaking it down into smaller, manageable subtasks. Recognizing her tendency toward procrastination, SRLAgent guides Elvira through structured goal-setting and effective time allocation, creating a realistic and achievable study plan. The integrated TimeHUD component visually represents her allocated time commitments for each subtask, helping Elvira maintain awareness and control over her schedule. All these preparation steps collectively form the Forethought Phase of SRL.

Next, Elvira moves into the active learning phase (Performance Phase). She engages with learning materials, which are gamified within Minecraft as interactive game items and entities to enhance motivation and engagement. As she progresses, the system continuously tracks her learning status and provides timely guidance through Large Language Model (LLM)-driven Non-Player Characters (NPCs). For instance, Elvira interacts with Knowledge Posters that visually represent complex concepts, takes quizzes to assess her understanding, and receives immediate feedback on challenging topics. When she encounters difficulties in understanding research papers, SRLAgent offers structured guidance to help her identify and summarize key insights. Additionally, Elvira interacts with Prof.NPC, a virtual professor powered by an LLM, who engages her in conversational question-answering sessions to clarify and resolve any lingering questions. To consolidate her understanding, Elvira completes a Report Writing Task, synthesizing and articulating what she has learned.

Upon completing her learning tasks, Elvira enters the Self-Reflection Phase. SRLAgent prompts her to thoughtfully consider several reflective questions: What key concepts have I learned? Which parts of the material were most challenging for me? Which learning strategies were effective or ineffective? How can I improve my approach to similar assignments in the future? This structured reflection process helps Elvira build critical metacognitive awareness, addressing gaps typically present in her study practices.

Throughout this learning journey, SRLAgent continuously monitors Elvira’s progress, providing adaptive and personalized guidance aligned with the development of her SRL skills.

\section{User Study}
To evaluate the effectiveness of SRLAgent in fostering student self-regulation, we conducted a between-subject user study, where students were randomly assigned to either use the SRLAgent or one of the two baselines to learn the artificial intelligence concept of the "Large Language Model (LLM) agent". Our research questions (RQs) are as follows:
\begin{itemize}
    \item \textbf{RQ1.} How does SRLAgent impact college students' self-regulated learning skills compared to traditional learning resources? 
    \item \textbf{RQ2.} How does SRL influence students’ academic performance?
    \item \textbf{RQ3.}  How do users perceive SRLAgent in terms of engagement and trust?
\end{itemize}


\subsection{Participants}
We recruited participants from a university in China, specifically targeting college freshmen who self-identified as requiring academic support and lacking effective study skills. These students had previously expressed interest in improving their study habits and academic outcomes. The final sample consisted of 45 participants (23 male and 22 female), with a mean age of 19 years (SD = 18.96). All participants are Chinese, and we screened for prior experience with LLM Agents to ensure participants have comparable baseline knowledge.  Participants were randomly assigned to either one of the three study condition. See details in ~\ref{chp:exp_set}.  All participants received 25 RMB for their participation in the one-hour study. 
\begin{table}[h]
\centering
\resizebox{0.48\textwidth}{!}{ 
\begin{tabular}{lcccc}
\toprule
& \textbf{Baseline A} & \textbf{System Group B1} & \textbf{System Group B2} & \textbf{Difference} \\

\midrule
\textbf{Gender} & & & & $\chi^2(2) = .297, p = .862$ \\
Female & 53.33\% ($N=8$) & 50\% ($N=7$) & 43.75\% ($N=7$) & \\
Male & 46.67\% ($N=7$) & 50\% ($N=7$) & 56.25\% ($N=9$) & \\
\midrule
\textbf{School} & & & & $\chi^2(8) = 2.856, p = .943$ \\
Humanities and Social Science & 13.33\% ($N=2$) & 7.14\% ($N=1$) & 12.50\% ($N=2$) & \\
Medicine, Life and Health Sciences & 20.00\% ($N=3$) & 28.57\% ($N=4$) & 12.50\% ($N=2$) & \\
Data Science & 33.33\% ($N=5$) & 28.57\% ($N=4$) & 25.00\% ($N=4$) & \\
Management and Economics & 26.67\% ($N=4$) & 14.29\% ($N=2$) & 25.00\% ($N=4$) & \\
Science and Engineering & 6.67\% ($N=1$) & 21.43\% ($N=3$) & 25.00\% ($N=4$) & \\
\midrule
\textbf{N} & 15 & 14 & 16 & \\
\bottomrule
\end{tabular}
}
\caption{Participant demographics and characteristics.}
\label{tab:participant_demographics}
\end{table}

\subsection{Experiment Setup} \label{chp:exp_set}
We employed a between-subjects experimental design to evaluate SRLAgent's effectiveness in enhancing self-regulated learning skills and learning outcomes. This approach allowed us to compare participants who used SRLAgent against those who used a baseline system, with pre- and post-test measurements to assess changes over time.

During recruitment, participants indicated their interest in learning about Large Language Models, specifically focusing on "LLM Agents" as the content domain. Each participant was randomly assigned to one of the following conditions:
\begin{itemize}
    \item \textbf{Baseline A (Multimedia Learning)}: Participants learned about learning materials through a pre-recorded instructional video, reflecting a common, traditional approach to online learning. This baseline is set to examine the effectiveness of SRLAgent compared to traditional learning method.
    \item \textbf{System Group B1 (SRLAgent without SRL features)}: Participants used a baseline version of SRLAgent that excluded all SRL-specific features, such as the goal-setting page, strategic planning page, LLM-assisted guidance, time-management tools, and self-reflection functionalities. The system still operated within the Minecraft environment, containing the same learning materials as the experimental version but without scaffolds for self-regulated learning. This group is set to investigate the effect of SRL features of SRLAgent on students' learning outcomes.
    \item \textbf{System Group B2 (SRLAgent)}: Participants used the complete SRLAgent system with all SRL features enabled.
\end{itemize}
To ensure consistency between different experiment conditions, both conditions maintained identical learning content and learning duration.


\begin{figure*}[htbp] 
    \centering 
    \includegraphics[width=0.9\textwidth]{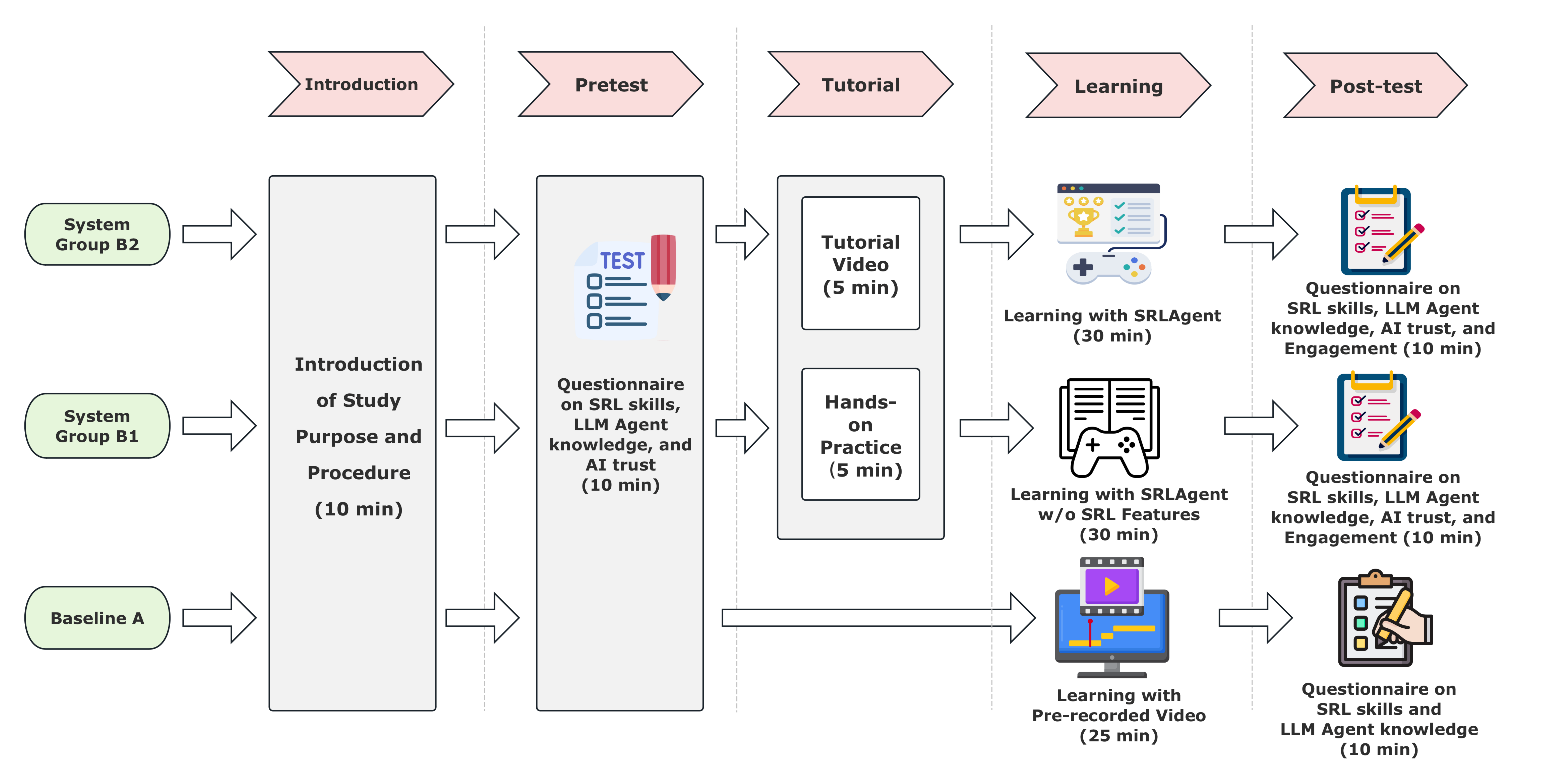}
    \caption{The procedure of the between-subject user study.}
    \label{fig:study_proc}
\end{figure*}

\subsection{Procedure}
The study was approved by the Institutional Review Board of the authors' institution. After obtaining informed consent, participants completed a five-phase protocol: introduction, pretest, tutorial, learning, and post-test. The complete procedure is illustrated in Figure~\ref{fig:study_proc}.
\subsubsection{\textit{Introduction (5 minutes)}}
The study was conducted in a computer laboratory at a university in China. Each participant was assigned to a computer equipped with a headset for audio input and output. A research assistant introduced the purpose of the study, provided an overview of the experimental procedure, and explained the tasks participants would be performing.
\subsubsection{Pretest (10 minutes)}
All participants completed identical pretest assessments measuring both self-regulated learning skills and prior knowledge about LLM agents. This served as a baseline for evaluating changes in their SRL skills and knowledge acquisition throughout the study. 

\subsubsection{Tutorial (10 minutes)}
Participants assigned to the system groups (B1 and B2) were given a tutorial session to familiarize themselves with the system’s features and the operations within the Minecraft environment. Participants in the multimedia learning group (baseline A) skipped this step and proceeded directly to the learning phase after the introduction session.
The tutorial was divided into two parts: a five-minute instructional video and a five-minute hands-on practice session. The instructional video provided an overview of the system interfaces and key features. While both groups watched similar videos, the content was tailored to their assigned condition—participants in the control group (B1) viewed a tutorial that only covered the baseline features, while those in the experimental group (B2) were introduced to the complete SRLAgent system, including all self-regulated learning components. After the video, participants completed a standardized navigation task to ensure they were comfortable using their assigned system before proceeding with the learning session.
\subsubsection{\textit{Learning (30 minutes)}} During this phase, participants in the system groups (B1 and B2) independently engaged with their assigned system to learn about LLM agents. The learning experience for participants in baseline A (multimedia learning) involved a 25-minute pre-recorded video that covered the same content. This baseline aimed to provide a traditional, non-interactive learning experience for comparison with the SRL-enhanced systems. Throughout the learning phase, research assistants were available for technical support but provided minimal interference to maintain experimental validity. Participants were encouraged to proceed at their own pace, ensuring that the learning experience was not disrupted by external guidance.
\subsubsection{\textit{Post-test (10 minutes)}}
Upon completion of the learning session, participants filled out a post-test questionnaire. This assessment aimed to measure any changes in SRL skills, knowledge of LLM agents, and perceived engagement with the system. The post-test also included questions to gauge participants' satisfaction with the learning experience and the system’s features. These results were compared to pretest responses to assess the effectiveness of the intervention in improving self-regulation skills and knowledge acquisition.

\subsection{Evaluation Metrics}
To evaluate the effectiveness of SRLAgent in fostering self-regulated learning (SRL) skills, we utilized a range of quantitative and qualitative measures to assess participants' SRL skills, learning outcomes, and engagement. These metrics provided insights into the system's impact across multiple dimensions, including changes in self-regulation abilities, knowledge acquisition, and overall engagement with the learning process.

\subsubsection{Self-Regulated Learning Skills (RQ1)}
Drawing from Nambiar's Academic Self-Regulated Learning Questionnaire (ASLQ)\cite{nambiar2022development}, the research team posed 18 questions to assess participants' self-regulated learning (SRL) skills. Example items include: "I don’t feel motivated to study difficult subjects", " I make sure that I complete the portions on time", and "I make necessary changes in my study plan to improve learning". Participants rated each item on a 7-point Likert scale (1 = strongly disagree, 7 = strongly agree). The final score, ranging from 1 to 7, was calculated by summing the points across all items and dividing the number of all items. The Cronbach’s alpha for this scale was 0.87, indicating good internal consistency. Full questionnaire details can be found in Appendix~\ref{tab:ASLQ}.

\subsubsection{Learning Outcomes (RQ2)}
To assess participants’ knowledge acquisition regarding the concept of LLM Agents, we adopted the 12-item official final test from the Hugging Face LLM Agent tutorial\footnote{\url{https://huggingface.co/learn/agents-course/unit1/final-quiz}}. Each item is a multiple-choice question that evaluates a key aspect of Large Language Models and agent usage, encompassing topics such as essential terminology and architecture. Participants earned 1 point for each correct answer, producing a total possible score of 12. This test was administered at both the pre- and post-test stages to gauge changes in participants’ knowledge. Higher scores indicate a stronger mastery of the LLM Agent content.

\subsubsection{Engagement (RQ3)}
To understand participants’ engagement in the activity, the research team analyzed their responses using a 30-item 5-point Likert scale (1-strongly disagree, 5-strongly agree) adapted from the User Engagement Scale (UES)~\cite{o2010development}. Example items include “I lost myself in this experience.”, “I felt frustrated while using this app.”, “This app was attractive”), and “Using app was worthwhile”. 
The scale was used in the post-test. Scale scores are calculated for each participant by summing scores for the items in each of the four subscales and dividing by the number of items. The overall engagement score can be calculated by adding the average of each subscale. The Cronbach’s alpha for this questionnaire was 0.90.

\subsubsection{Trust in AI System (RQ3)}
We used a 12-item questionnaire adapted from Jian’s Trust Scale~\cite{jian2000foundations} to measure participants’ trust in the AI system before and after the intervention. This scale, validated in AI-related contexts~\cite{scharowski2025trustdistrusttrustmeasures}, captures both trust and distrust through statements like “The System is deceptive”, “I am wary of the System”, and “I can trust the System”. Participants indicated their level of agreement on a 7-point Likert scale ranging from 1 (“strongly disagree”) to 7 (“strongly agree”). Certain items were reverse-coded to account for negative statements. An overall trust score was calculated by averaging all item responses, with higher values indicating higher levels of trust in SRLAgent. The Cronbach’s alpha exceeded 0.80, suggesting a good to excellent level of internal consistency for the scale.


\section{Results}


\subsection{Self-Regulated Learning Skills}

To evaluate the effectiveness of SRLAgent in enhancing participants' self-regulated learning (SRL) skills, we analyzed changes in SRL scores.

\begin{figure*}[htbp] 
\centering
\includegraphics[width=0.9\textwidth]{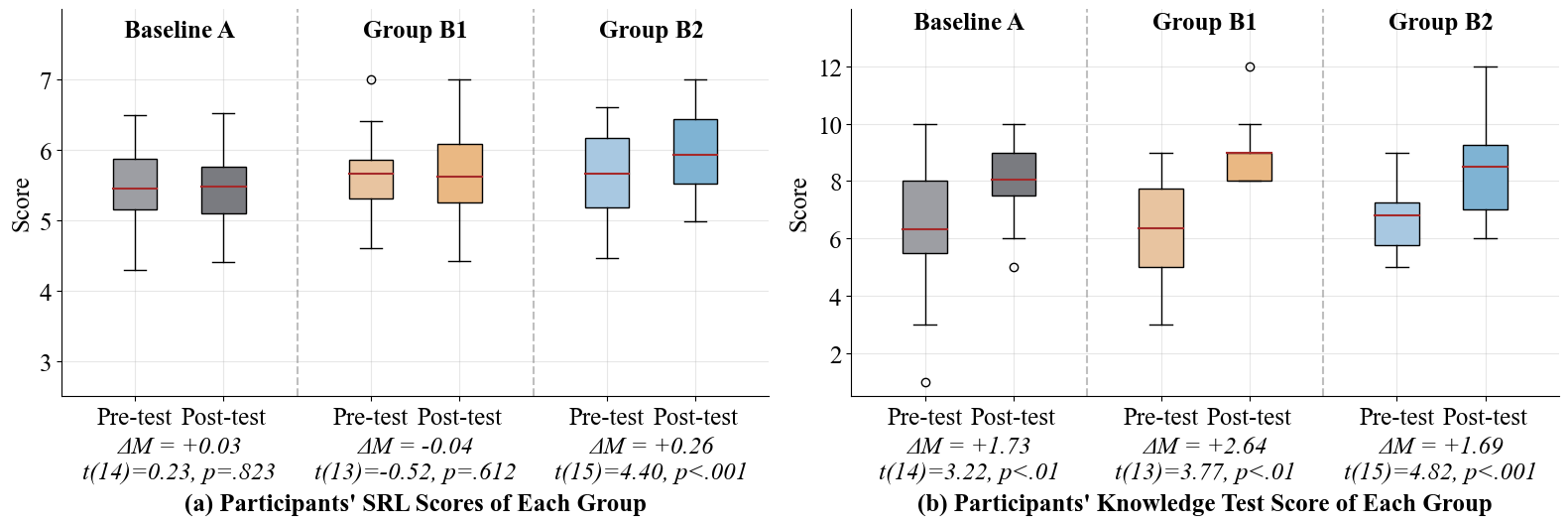}
\caption{Distribution of participants' (a) self-regulated learning (SRL) scores and (b) LLM agent knowledge test scores before and after the intervention for the SRLAgent (Group B2), SRLAgent without SRL features (Group B1), and Multimedia Learning groups (Baseline A). Statistically significant results are indicated as $p$ < .05*, $p$ < .01**, $p$ < .001***.}
\label{fig: srl&quiz_score}
\end{figure*}

We conducted paired-samples t-tests to assess changes in self-regulated learning (SRL) skills within each experimental condition (SRLAgent, SRLAgent without SRL features, and Multimedia Learning). As shown in Figure~\ref{fig: srl&quiz_score} (a), participants who interacted with the complete SRLAgent showed a statistically significant improvement in their SRL scores from pretest (M = 5.66, SD = .67) to post-test (M = 5.92, SD = .65), with \textit{t}(15) = 4.41, \textit{p} < .001, Cohen’s \textit{d} = .234. This indicates that the integrated SRL scaffolds provided by SRLAgent effectively enhanced participants' overall self-regulation abilities. Conversely, no significant improvements from pretest to post-test were observed in either the SRLAgent without SRL features group (\textit{t}(13) = .15, \textit{p} = .883) or the Multimedia Learning group (\textit{t}(14) = .24, \textit{p} = .814). These results suggest that neither traditional multimedia instruction nor the Minecraft-based platform alone substantially impacted participants' SRL skills, highlighting the importance of specific SRL-focused design features.

To assess baseline equivalence, a one-way ANOVA was performed on pre-test SRL scores across the three groups (SRLAgent, SRLAgent without SRL features, and Multimedia Learning). The analysis revealed no significant differences, F(2, 42) = .69, \textit{p} = .507, confirming comparable starting levels. An independent-samples t-test comparing post-test SRL scores between SRLAgent and Multimedia Learning groups showed a marginally significant difference, \textit{t}(29) = 1.93, \textit{p} = .063, Cohen’s \textit{d} = .638, indicating a moderate effect size favoring SRLAgent.

\begin{figure*}[htbp] 
\centering
\includegraphics[width=0.9\textwidth]{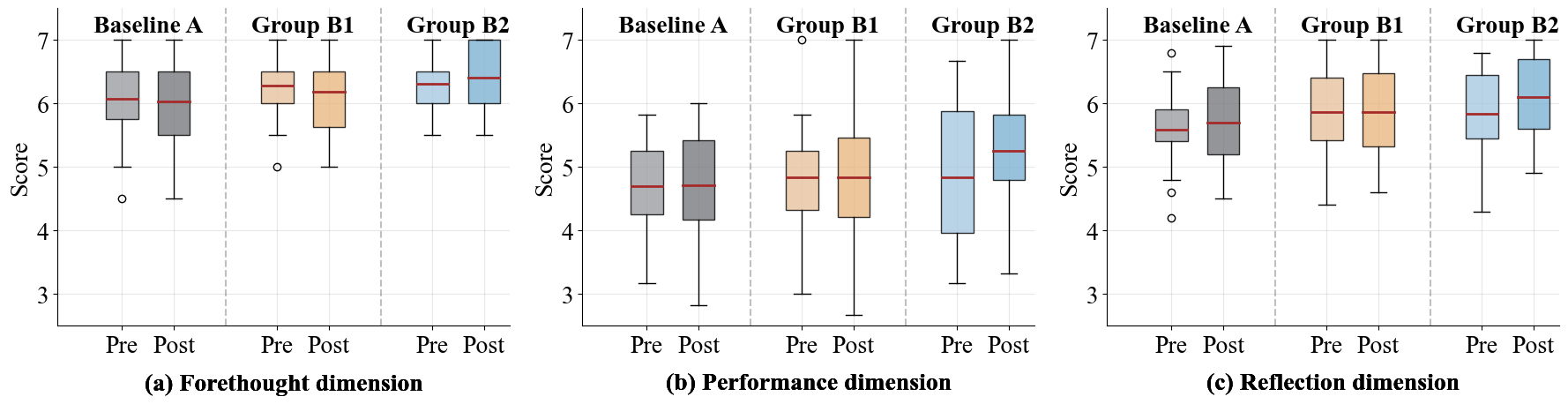}
\caption{Distribution of participants' score on (a) Forethought, (b) Performance, and (c) Reflection dimensions of ASLQ before and after the intervention for the SRLAgent (Group B2), SRLAgent without SRL features (Group B1), and Multimedia Learning groups (Baseline A).}
\label{fig: srl_each_stage_score}
\end{figure*}

\subsection{Learning Outcomes}
To examine learning outcomes across the three intervention conditions—SRLAgent, SRLAgent without SRL features, and Multimedia Learning—we compared participants’ pre- and post-test performance scores. As shown in Figure~\ref{fig: srl&quiz_score} (b), descriptive statistics indicated gains in all groups, with the SRLAgent without SRL features group showing the largest improvement (M diff = 2.53, SD = 2.56), followed by the Multimedia Learning group (M diff = 1.73, SD = 2.08) and the SRLAgent group (M diff = 1.69, SD = 1.40). To further compare post-intervention performance, an independent-samples t-test was conducted between the SRLAgent and SRLAgent without SRL features groups, and the analysis revealed no statistically significant difference, \textit{t}(28) = 1.59, \textit{p} = .123.


\subsection{Engagement}
To analyze user engagement across the SRLAgent and baseline version of SRLAgent (without SRL features), we conducted an independent-samples t-test to examine differences in learner engagement between participants of two systems. Although no statistically significant difference emerged (\textit{t}(28) = 1.59, \textit{p} = .123), participants in the SRLAgent group reported higher average engagement scores (M = 4.17, SD = .45) compared to the group without SRL features (M = 3.80, SD = .82). This descriptive trend suggests that integrating SRL scaffolding and adaptive feedback provided by a large language model may positively influence learner engagement. Further research with larger samples may help confirm and clarify these preliminary findings.

\begin{figure}[htbp] 
\centering
\includegraphics[width=0.5\textwidth]{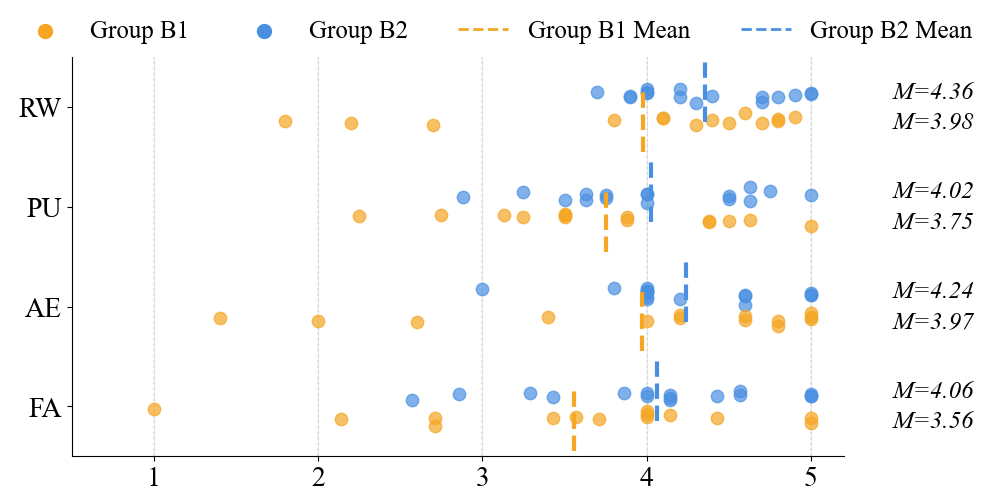}
\caption{Scatter plot illustrating the distribution of participants' engagement level across four engagement dimensions (Focused Attention (FA), Perceived Usability (PA), Aesthetic Elements (AE), Reward Factor (RW)) for SRLAgent group (Group B2) and SRLAgent without SRL features group (Group B1). }
\label{fig: engagement_score}
\end{figure}

\subsection{Trust in AI}
To examine whether using SRLAgent influenced participants' trust in AI, we conducted a paired-samples t-test comparing pre- and post-intervention trust scores within the SRLAgent group. Participants exhibited a modest increase in trust scores from the pretest (M = 4.66, SD = .73) to the post-test (M = 4.91, SD = .80). However, this improvement was not statistically significant, \textit{t}(15) = 1.62, \textit{p} = .127. Despite the absence of statistical significance, the observed increase in the SRLAgent group was notably larger compared to negligible changes found in the Multimedia Learning ($\Delta M \approx 0$) and SRLAgent without SRL features groups ($\Delta M$ = .006; see Figure X).

We further conducted a one-way ANOVA comparing post-test trust scores across the three conditions (SRLAgent, SRLAgent without SRL features, and Multimedia Learning). The results did not yield a statistically significant effect of intervention type, F(2, 42) = 1.21, \textit{p} = .310. Nevertheless, descriptive statistics revealed a consistent trend favoring SRLAgent, with participants reporting higher post-test trust scores (M = 4.91, SD = .80) relative to the Multimedia Learning condition (M = 4.62, SD = .80) and SRLAgent without SRL features (M = 4.52, SD = .60).

Although the results lack statistical significance—potentially due to the relatively small sample size—they highlight a meaningful descriptive trend suggesting that enhanced SRL scaffolding and personalized AI-driven feedback may positively influence students' trust in AI-based learning platforms.

\section{Discussion}

\subsection{Fostering Self-Regulated Learning Skills and Engagement with SRLAgent}

The findings demonstrate the potential of SRLAgent in effectively promoting college students' self-regulated learning (SRL) skills. The significant improvements observed support earlier studies suggesting that structured planning, feedback, and scaffolding can positively influence students' abilities to regulate their learning processes \citep{schunk2011handbook, zimmerman2002becoming}. 

Although our comparative analyses across conditions did not yield statistically significant group differences, the consistently higher SRL scores among students using SRLAgent suggest practical benefits of incorporating comprehensive SRL scaffolding into learning technologies. This finding resonates with prior work indicating that embedding self-regulatory support within digital platforms can benefit learners \citep{broadbent2015self, bellhauser2016applying}.

Regarding engagement, the observed higher mean engagement scores indicate that SRLAgent's design has a promising potential to maintain learners' attention and motivation. This aligns with existing research identifying the motivational benefits of adaptive feedback and interactive gamification elements in educational systems \citep{landers2014developing, koivisto2019rise}. Specifically, the gamified, quest-based structure and interactive feedback provided by SRLAgent likely contribute to sustaining learners' intrinsic motivation and promoting deeper immersion in the learning task.

In summary, SRLAgent fosters SRL among college students while maintaining learner engagement. These insights both confirm and extend existing research on SRL-supportive educational technologies and highlight specific opportunities for future enhancements to better scaffold learners throughout the complete SRL process.



\subsection{Enhancing Academic Performance with SRL}

Our findings suggest that SRLAgent had a positive effect on academic performance, as evidenced by the significant improvement in participants' understanding of the Large Language Model (LLM) concept. This aligns with previous studies demonstrating that self-regulated learning interventions can lead to better knowledge acquisition and retention~\cite{paris2003classroom}. The ability of SRLAgent to provide personalized, adaptive feedback likely contributed to this improvement, as students received tailored guidance that aligned with their individual learning needs.

Further research could explore how specific elements of SRLAgent, such as the pacing of feedback or the integration of task management tools, influence academic performance. Additionally, measuring long-term retention would be important to assess whether the benefits of SRLAgent are sustained beyond the immediate learning period.

\subsection{Balancing between Autonomous and AI Assistance in the Learning Process}

The results revealed an interesting trade-off between learning outcomes and trust in AI assistance. Specifically, the SRLAgent group, which received comprehensive AI-driven SRL features, showed lower improvement in learning outcome scores from pretest to post-test compared to the SRLAgent without SRL features group. Conversely, participants in the SRLAgent group reported higher trust scores toward the AI system. This discrepancy suggests that while AI assistance can enhance user trust and perceived system reliability, excessive reliance on AI support might inadvertently reduce students' opportunities for independent thinking, problem-solving, and deeper cognitive engagement.

Previous research argues that overly scaffolded learning environments may lead learners to adopt passive learning behaviors, hindering the development of self-regulated learning skills and critical thinking abilities \cite{koedinger2012knowledge}. Our findings align with this perspective, indicating that participants who received substantial AI support might have become overly dependent on the system's guidance, thus limiting their autonomous exploration and cognitive effort during the learning process.

To address this issue, future iterations of SRLAgent could implement adaptive fading strategies, gradually reducing AI support as learners demonstrate proficiency and confidence in targeted skills. For instance, the system first provides stronger initial scaffolding and progressively reduces guidance, fostering learners to progressively assume greater responsibility for their learning and engage deeper in the cognitive process of learning \cite{pekrun2006control}. Additionally, incorporating Chain-of-Thought (CoT) or step-by-step interactive questioning techniques could promote deeper conceptual understanding. Such interactive questioning encourages learners to actively articulate their reasoning processes, thereby fostering more meaningful engagement with the learning material and reducing reliance on AI-generated answers.

\subsection{Limitation and Future Work}
Our study provides initial insights into how SRLAgent can support college students' self-regulated learning (SRL) skills, engagement, and trust in AI-driven educational environments. Nevertheless, several limitations warrant consideration and highlight important avenues for future research.

First, our relatively small sample size limited statistical power, potentially constraining the detection of significant differences, especially in measures of engagement and trust. Although descriptive trends indicated positive outcomes, future studies with larger samples would help clarify these preliminary findings and allow for more robust statistical conclusions.

Second, the short-term nature of our intervention (a single learning session) might have restricted observable changes. Longitudinal studies are necessary to investigate sustained effects over time and to understand how prolonged interaction with SRLAgent might influence deeper reflective practices, long-term engagement, and evolving perceptions of AI trustworthiness.

Third, our measures of trust and engagement relied primarily on quantitative self-report scales, which may not fully capture nuanced user perceptions and behaviors. Integrating qualitative approaches, such as interviews, think-aloud protocols, or behavioral analytics, would provide richer insights into how students perceive and interact with educational AI systems, contributing to a more comprehensive understanding of learner experiences.

Finally, our study focused on college students within a specific learning context (understanding LLM agents). Future research should examine SRLAgent’s applicability across diverse educational settings, subjects, and learner demographics to assess its generalizability and adaptability. Exploring how different student populations engage with and benefit from SRL scaffolding and LLM-based feedback could yield valuable implications for inclusive and personalized educational technology design.

In summary, addressing these limitations in future research will improve our understanding of SRL-supportive AI systems, inform the design of more effective educational technologies, and ultimately contribute to richer, more engaging, and trustworthy learning experiences.

\section{Conclusion}
In this study, we designed and evaluated SRLAgent, an AI-powered educational system that integrates self-regulated learning (SRL) scaffolding and personalized feedback to improve learners' SRL skills, engagement, and trust toward AI-supported learning environments. Our findings demonstrate that SRLAgent effectively promotes learners' abilities in the forethought and performance stages of Zimmerman's SRL model, highlighting the value of explicit goal-setting and real-time feedback mechanisms. Although improvements in the self-reflection dimension were limited, these results underscore the importance of comprehensive SRL support, including structured reflective activities, to fully realize the potential of educational AI technologies.

Furthermore, descriptive trends indicate that SRLAgent may positively influence learner engagement and trust in AI systems, suggesting promising directions for future research and design improvements. However, the absence of statistically significant findings in these areas points to the need for larger-scale studies and more sensitive measures to capture nuanced behavioral and attitudinal changes.

Overall, our work contributes valuable insights into how carefully designed SRL features and adaptive AI feedback can assist learners in developing essential regulatory skills and maintaining motivation. Future research should build upon these findings by exploring robust reflection supports, expanding participant samples, and incorporating multimodal data collection methods. Addressing these directions will advance our understanding of effective AI-assisted learning designs, ultimately fostering deeper and more sustained educational outcomes.

\clearpage

\clearpage

\bibliographystyle{ACM-Reference-Format}
\bibliography{Files/ref}

\clearpage

\appendix
\section{Appendix}
\subsection{Prompt Configurations for LLM Agents}
In this section, we illustrate the diverse prompts sent to each agent in SRLAgent.

\begin{figure*}[!t]
\centering
\footnotesize
    \begin{tcolorbox}[aibox, title=Prompt Design for Planning Agent, width=0.9\textwidth]
    \textbf{System: } You are a Task Planning Assistant specializing in Self-Regulated Learning strategies. Your role is to help students develop effective task planning skills by analyzing subtasks, suggesting optimal ordering, and providing strategic guidance. Format your responses with clear structure and reasoning to model effective planning processes.\\
    
    \{chatHistory\}\\
    
    \textbf{User: } 
    \# Task Planning Request
    
    I'm working on the Self-Regulated Learning (SRL) task planning phase. Please help me organize the following subtasks efficiently while developing my SRL thinking.
    
    \#\# Considerations:
    * Task dependencies and logical sequence
    * Resource utilization efficiency
    * Time and difficulty balance
    
    \#\# Subtask List:
    1. \{Subtask title\}
       * Description: \{Subtask description\}
       * Estimated time: \{Time estimate\}
    
    2. \{Additional subtasks...\}
    
    \#\# Response Format Requirements:
    1. \*\*Optimal Sequence:\*\*
       \texttt{\\begin\{verbatim\}}
       <START>
       Comma-separated task numbers in optimal order (e.g., 3,1,5,2,4)
       <END>
       \texttt{\\end\{verbatim\}}
    
    2. \*\*Reasoning:\*\*
       Explain the rationale behind this sequence
    
    3. \*\*Completion Strategy:\*\*
       Provide recommendations for effective task execution
    
    \*Note: The <START> and <END> tags are required for automated processing.\*
    \end{tcolorbox}
    \captionof{figure}{Planning Agent Prompt Configuration}
    \label{fig:planning-agent-prompt}

\end{figure*}

    
    
    
    
    
    

\begin{figure}[htpb]
    \begin{tcolorbox}[aibox, title=Prompt Design for Reflection Agent, width=0.45\textwidth]
    \textbf{System: } You are a Reflection Agent specializing in Self-Regulated Learning (SRL) strategies. Your role is to help learners develop effective reflection skills by analyzing their task performance, identifying strengths and areas for improvement, and guiding them through meaningful self-evaluation.
    
    When providing reflection guidance:
    1. Focus on specific task outcomes and SRL strategies used
    2. Highlight connections between planning, execution, and results
    3. Encourage metacognitive thinking about learning processes
    4. Suggest actionable improvements for future learning tasks
    
    Keep your feedback concise, constructive, and clearly structured. \\

    \{chatHistory\} \\

    \textbf{User: } 
    \# Reflection Request
    
    I'm in the reflection phase of my Self-Regulated Learning (SRL) process. Please help me analyze my task performance and generate meaningful insights.
    
    \#\# Task Information:
    * Task Title: \{Task title\}
    * Task Description: \{Task description\}
    
    \#\# Subtask Completion Details:
    
    \{Subtask outcomes will be listed here\}
    
    \#\# Reflection Requirements:
    1. \*\*Performance Summary:\*\* (30 words maximum)
       Provide a concise assessment of task execution
    
    2. \*\*SRL Strategy Analysis:\*\*
       Identify which self-regulated learning strategies were effectively applied
    
    3. \*\*Learning Insights:\*\*
       Highlight key takeaways and potential improvements for future tasks
    
    Please focus on both the task outcomes and the learning process itself, helping me develop stronger self-regulation skills.
    \end{tcolorbox}
    \caption{Reflection Agent Prompt Configuration}
    \label{fig:reflection-agent-prompt}
\end{figure}

\begin{figure}[htpb]
    \begin{tcolorbox}[aibox, title=Prompt Design for Quiz Tutor Agent, width=0.45\textwidth]
    \textbf{System: } You are an educational support agent specializing in university-level academic content. Your role is to provide concise, helpful guidance when students encounter difficulties with different question types.
    
    The student is currently in the Performance phase of Self-Regulated Learning (SRL), working on knowledge acquisition subtasks. Your support should reinforce effective learning strategies while providing just enough guidance to overcome obstacles.
    
    Keep explanations brief (under 20 words) while maintaining clarity and educational value. Focus on promoting understanding rather than simply providing answers.\\

    \{chatHistory\}\\

    \textbf{User: } 
    \# Question Support Request
    
    I need help with \{\textbf{question type}\}.
    
    \{question details\}
    
    Please provide targeted guidance that helps me understand the conceptual relationships without giving away complete answers. Focus on promoting learning rather than simply providing solutions.
    
    \{condition \{\textbf{question details}\} based on \textbf{{question type}}:
    
    If \textbf{matching question}:
    "Based on the concept category \{concept category\} and the {incorrect connections} I've made, please provide a targeted hint."

    If \textbf{multiple choice}:
    "Explain why \"\{correct option\}\" is the correct definition of \{concept name\}"
    
    If \textbf{ordering/sequencing}:
    "Hint: \{current item at first error position\} should be placed at a different position in the \{ordering topic\} timeline."
    
    If \textbf{true/false}:
    "Briefly explain why the statement \{statement\} is \{true/false judgment\}"
    \}
    \end{tcolorbox}
    \caption{Quiz Tutor Agent Prompt Configuration}
    \label{fig:question-agent-prompt}
\end{figure}

\begin{figure}[htpb]
    \begin{tcolorbox}[aibox, title=Prompt Design for Chatting Agent, width=0.45\textwidth]
    \textbf{System: } You are Professor \{professor name\} from the \{department\} at \{university\}, a leading expert in the field of \{research field\}.
    
    Your research interests in \{specific area\} include \{research direction 1\}, \{research direction 2\},...
    
    You provide academic guidance in your field, including research directions, paper writing, and experiment design.
    
    You are welcoming students and answering their questions. \\

    \{chatHistory\}\\

    \textbf{User: } 
    
    \{user question\}
    \end{tcolorbox}
    \caption{Chatting Agent Prompt Configuration}
    \label{fig:professor-advisor-prompt}
\end{figure}

\begin{figure}[htpb]
    \begin{tcolorbox}[aibox, title=Prompt Design for Paper Review Agent, width=0.45\textwidth]
    \textbf{System: } You are an academic review expert. Please help players summarize the given paper abstract.
    
    Players are performing the execution step in a self-regulated learning (SRL) strategy, and the abstract is the planning step.
    
    Use critical thinking to help players understand the research ideas, core contributions, highlights, advantages and disadvantages of the paper.
    
    No more than 30 words.
    
    Please be concise, constructive, and clearly structured.\\
    
    \{chatHistory\}\\

    \textbf{User: } 
    \{combinedInput\}
    
    Where combinedInput may include:
    Question: \{question\}
    Summary: \{summary\}
    Paper Content:
    \{paperContent\}
    \end{tcolorbox}
    \caption{Paper Review Agent Prompt Configuration}
    \label{fig:paper-review-agent-prompt}
\end{figure}

\begin{figure}[t]
    \begin{tcolorbox}[aibox, title=Prompt Design forPaper Writing Agent, width=0.45\textwidth]
    \textbf{System: } You are an expert in adaptive paper writing. Your role is to assist the player in improving their adaptive writing skills, including structure, organization, and clarity.
    
    Focus on the adaptive writing process and critical thinking strategies. Provide step-by-step guidance and examples where necessary.

    Reply with no more than 50 words. Be concise and structured.
    
    Pay attention to proper citation and referencing formats.\\

    \{chatHistory\}\\

    \textbf{User: } 
    \{referenceContent\}
    
    Title: \{title\}
    Body: \{body\}
    Question: \{question\}
    
    \{reference content structure:
    If available, referenceContent will be formatted as:
    
    "Previous Task Outcomes:
    [Previous work and outcomes from related subtasks that should inform the current writing task. May include research findings, literature reviews, methodology descriptions, or other academic content the player has already produced.]"
    \}
    \end{tcolorbox}
    \caption{Paper Writing Agent Prompt Configuration}
    \label{fig:paper-writing-agent-prompt}
\end{figure}

\begin{table*}[!h]
\centering
\caption{Questionnaire for SRL skills}
\label{tab:ASLQ}
\begin{tabular}{p{0.5cm} p{13.5cm}}
\hline
\textbf{No.} & \textbf{Statement} \\
\hline
1 & I study in a suitable place where I can concentrate. \\
2 & When I am reading, I stop once in a while to review what I have read. \\
3 & I make necessary changes in study plan to improve learning. \\
4 & I don't feel motivated to study difficult subjects. \\
5 & I split my portions while studying. \\
6 & I go through the study material carefully to understand it properly. \\
7 & Before I start studying, I make a schedule. \\
8 & I try to strengthen the strategies that worked for me previously. \\
9 & I study in a manner that makes it more interesting/enjoyable. \\
10 & I use keywords/abbreviations to improve learning. \\
11 & When my studies are affected, I try to identify my mistakes. \\
12 & I learn by teaching others. \\
13 & I set targets before I start studying. \\
14 & While I am studying, I try to get rid of any distractions that are around me. \\
15 & I keep track of study areas where I am lacking. \\
16 & I don't have the habit of maintaining notes. \\
17 & I organize the study material before I start studying. \\
18 & After my exam I reflect upon areas I could have done better. \\
19 & I make notes to simplify learning. \\
20 & I try to learn from the mistakes I made in the exam. \\
21 & I constantly assess the amount of effort I put into my studies. \\
22 & I memorize key words to remind me of important concepts. \\
23 & Before I study, I make an outline of the content. \\
24 & I focus more on difficult portions while studying. \\
25 & I organize my time according to the difficulty of the task. \\
26 & I make sure that I complete the portions on time. \\
27 & If I miss a class, I take the help of others to cover the portions. \\
28 & I keep my assignments and class notes complete. \\
29 & I motivate myself to do better than before. \\
30 & While studying, I utilize different sources of information (lectures, reading, and discussions). \\
31 & I set a goal for how much to study each day. \\
32 & I make simple charts, diagrams, or tables while studying. \\
33 & I seek help when unable to understand a concept. \\
34 & When I study, I try to understand the concepts. \\
35 & I refer to my class notes whenever necessary. \\
36 & I make sure that I attend class regularly. \\
\hline
\end{tabular}
\end{table*}

\begin{table*}[!h]
\centering
\caption{AI Trust Questionnaire Items}
\label{tab:ai_trust_questionnaire}
\begin{tabular}{p{0.5cm} p{13.5cm}}
\hline
\textbf{No.} & \textbf{Statement} \\
\hline
1  & The AI is deceptive (R) \\
2  & The AI behaves in an underhanded manner (R) \\
3  & I am suspicious of the AI’s intent, action, or outputs (R) \\
4  & I am wary of the AI (R) \\
5  & The AI’s actions will have a harmful or injurious outcome (R) \\
6  & I am confident in the AI \\
7  & The AI provides security \\
8  & The AI has integrity \\
9  & The AI is dependable \\
10 & The AI is reliable \\
11 & I can trust the AI \\
12 & I am familiar with the AI \\
\hline
\end{tabular}
\end{table*}

\end{document}